\documentclass[utf8,latin1,aps,prl,showpacs,twocolumn,floats,superscriptaddress,10pt]{revtex4-1}
\usepackage{dcolumn}
\usepackage[pdftex]{graphicx}

\usepackage[UKenglish]{babel}
\usepackage{amsmath,amssymb,euscript}
\usepackage{mathtools}
\usepackage{placeins}

\usepackage{ifthen}
\newcommand{\hideandshow}[1]{%
 \ifthenelse{\isundefined{\showme}}{}{#1}}
\newcommand{\showandhide}[1]{%
 \ifthenelse{\isdefined{\showme}}{}{#1}}

\RequirePackage[
   hyperindex,colorlinks,bookmarksnumbered,
   plainpages=true,pdfstartview=FitH]{hyperref}
\hypersetup{linkcolor=blue,urlcolor=blue,citecolor=blue}
\usepackage{hyperref}

\usepackage{color}
\usepackage{ulem}
\renewcommand{\emph}[1]{\textit{#1}}
\definecolor{darkblue}{rgb}{0,0,0.5}
\definecolor{darkgreen}{rgb}{0,0.5,0}
\definecolor{darkred}{rgb}{.7,0,0}
\definecolor{purple}{rgb}{0.5,0,0.6}
\definecolor{orange}{rgb}{1,0.5,0}
\definecolor{grey}{rgb}{.6,.6,.6}
\definecolor{lightpink}{rgb}{1,0.7,0.75}
\definecolor{pink}{rgb}{1,0.4,0.58}
\definecolor{deeppink}{rgb}{1,0.08,0.58}

\newcommand{\scrap}[1]{}
\newcommand{\todo}[1]{}

\newcommand{\tocite}[1]{}
\newcommand{\comment}[1]{}
\newcommand{\question}[1]{}

\newcommand{\AHM}{{\rm AHM}}

\newcommand{\bath}{{\rm bath}}
\newcommand{\interact}{{\rm int}}
\newcommand{\charge}{{\rm ch}}
\newcommand{\spin}{{\rm sp}}
\newcommand{\orb}{{\rm orb}}

\newcommand{\pdag}{{\phantom{\dagger}}}
\newcommand{\Tk}{T_{\rm K}}
\newcommand{\Tkspin}{T_{\rm K}^\spin}
\newcommand{\Tkorb}{T_{\rm K}^\orb}
\newcommand{\mi}{{\rm i}}

\begin{document}

\title{DMFT+NRG study of spin-orbital separation in a three-band
Hund's metal}

\author{K. M. Stadler} \affiliation{Physics
  Department, Arnold Sommerfeld Center for Theoretical Physics and
  Center for NanoScience, Ludwig-Maximilians-Universit\"at M\"unchen,
  80333 M\"unchen, Germany} 
 \author{A. Weichselbaum} \affiliation{Physics
  Department, Arnold Sommerfeld Center for Theoretical Physics and
  Center for NanoScience, Ludwig-Maximilians-Universit\"at M\"unchen,
  80333 M\"unchen, Germany}  
  \author{Z. P. Yin} \affiliation{Department of Physics and Astronomy, Rutgers University, Piscataway, NJ 08854, USA} 
\author{J. von Delft} \affiliation{Physics
  Department, Arnold Sommerfeld Center for Theoretical Physics and
  Center for NanoScience, Ludwig-Maximilians-Universit\"at M\"unchen,
  80333 M\"unchen, Germany}
  \author{G. Kotliar} \affiliation{Department of Physics and Astronomy, Rutgers University, Piscataway, NJ 08854, USA} 
\date{March 22, 2015}
\begin{abstract}
  \date{\today} We show that the numerical renormalization group (NRG)
  is a viable multi-band impurity solver for Dynamical Mean Field
  Theory (DMFT), offering unprecedent real-frequency spectral
  resolution at arbitrarily low energies and temperatures.  We use it
  to obtain a numerically exact DMFT solution to the Hund's metal
  problem for a three-orbital model with filling factor $n_d=2$. The
  ground state is a Fermi liquid. The one-particle spectral function
  shows a coherence-incoherence crossover with increasing temperature,
  with spectral weight being transfered from low to high energies, and
  evolves qualitatively differently from a doped Mott insulator.  The
  spectral function has a strong particle-hole asymmetry, and in the
  incoherent regime the one-particle self-energy shows approximate
  power-law behavior for positive frequencies only.  The spin and
  orbital spectral functions show ``spin-orbital separation'': spin
  screening occurs at much lower energies than orbital screening.  The
  renormalization group flows clearly reveal the relevant physics at
  all energy scales.
\end{abstract}
\pacs{71.27.+a, 71.10.Fd, 75.20.Hr\vspace{-5mm}}

\maketitle


\paragraph*{Introduction.---} 
\label{sec:intro}

A widely-used method for dealing with interactions in
strongly-correlated electron systems and electronic structure
calculations is dynamical mean field theory (DMFT)
\cite{Georges1996,Kotliar2006}.  It treats the interplay be\-tween a
given lattice site (the ``impurity'') and the rest of the lattice (the
``bath'') as a quantum impurity model with a self-consistently
determined hybridization function.  Since DMFT's performance depends
on that of the method used to solve this impurity model, much effort
has been invested over the years to develop ever more powerful
impurity solvers.  For multi-band models, con\-ti\-nuous-time Quantum
Monte Carlo (ctQMC) methods appear to be the current favorites in
terms of versatility and performance \cite{Gull2011a}.  However, they
are not without limitations: sign problems can occur, low-temperature
calculations are costly, and obtaining real-frequency spectra requires
analytic continuation of imaginary (Matsubara) frequency QMC data,
which is notoriously difficult. Thus, there is a continued need for
\textit{real-frequency} impurity solvers suitable for
\textit{multi-band} DMFT applications.

In this work, we show that the numerical renormalization group (NRG)
\cite{Wilson1975,Bulla2008,Weichselbaum2012a} is such a tool, offering
unprecedented real-frequency spectral resolution at low energies. NRG
is the gold standard for impurity models, with numerous previous DMFT
applications (e.g.\ \cite{Sakai1994,Bulla1998,Bulla1999,%
  Bulla2001,Pruschke2005,Deng2013,Osolin2015}), but until recently was
limited to models with at most two bands. However, recent
technical progress \cite{Toth2008,Weichselbaum2012b,Mitchell2014}
has now made three-band calculations feasible
\cite{Moca2012,Hanl2013,Hanl2014}.

We illustrate the potential of DMFT+NRG by studying a three-band model
\cite{Yin2012,Georges2013,Aron2015} with both a Hubbard
interaction $U$ and a ferromagnetic Hund's coupling $J$, with
$\text{U(1)}_\charge\times%
\text{SU(2)}_\spin \times\text{SU(3)}_\orb$ symmetry for its charge
(\charge), spin (\spin) and orbital (\orb) degrees of freedom.
This is the simplest model of a 3-band ``Hund's metal''
\cite{Haule2009,Yin2011}. These are multi-orbital materials with broad
bands which are correlated via the Hund-$J$ rather than the
Hubbard-$U$ interaction. Examples are iron pnictide and chalcogenide
high-temperature superconductors \cite{Haule2009,Yin2011a}, 
ruthenates \cite{Werner2008,Mravlje2011}, and other 4d transition metal oxides
\cite{Medici2011,Georges2013}.

Early DMFT studies using continuous-time QMC (ctQMC) \cite{Gull2011a}
as impurity solver suggest that consequences of the Hund's rule
coupling include (i) Fermi-liquid behavior at low energies
\cite{Haule2009} and (ii) a coherence-incoherence crossover with
increasing temperature \cite{Haule2009}, relevant for various material
systems \cite{Mravlje2011,Hardy2013}. The incoherent regime is
characterized by (iii) fractional power laws
\cite{Werner2008,Yin2012,Akhanjee2013} and (iv) the coexistence of
fast quantum mechanical orbital fluctuations and slow spin
fluctuations \cite{Yin2012}. However, since ctQMC can not reach truly
low temperatures, (i) could not be conclusively established yet, and a
more detailed understanding of (ii-iv) is difficult to achieve based
on imaginary-frequency data alone. Our real-frequency DMFT+NRG results
definitively settle these issues and yield further insights.  We find
(i) a Fermi liquid ground state; a real-frequency one-particle
spectral function showing (ii) a coherence-decoherence crossover with
significant transfer of spectral weight from low to high energies, and
(iii) strong particle-hole asymmetry, which leads to the
above-mentioned apparent fractional power laws; and (iv) two-stage
screening, where spin screening occurs at much lower energies than
orbital screening (``spin-orbital separation'').  (v) The $T=0$
spectral properties are similar with or without DMFT self-consistency,
in marked contrast to Mott-Hubbard systems, where the DMFT
self-consistency opens a gap in the quasiparticle spectrum at large
interaction strength.

\paragraph*{Model:---}
\label{sec_Models}
Our three-band model has Hamiltonian
\begin{subequations}
\label{eq:HU-Hloc}
\begin{eqnarray}
  \hat{H} & = & \sum_{i} \left(  - \mu \hat N_{i} 
+ \hat{H}_\interact [\hat d^\dag_{i\nu}] \right)  \label{eq:HU}
+\sum_{\langle ij\rangle \nu} t\,
  \hat{d}^{\dagger}_{i\nu}\hat{d}^{\phantom{\dagger}}_{j\nu}  , 
\\
\hat{H}_\interact[\hat d^\dag_{i\nu} ]  \label{eq:Hloc}
& = & \tfrac{3}{4} {J}  \hat N_i +
\tfrac{1}{2}\left( U-\tfrac{1}{2}J \right) 
\hat N_i (\hat N_i -1)-{J}\hat{\mathbf S}_i^2. \qquad
\end{eqnarray}
\end{subequations}
Here $\hat d^\dagger_{i \nu}$ creates an electron on site $i$ of
flavor (fl) $\nu$, with composite index $\nu = (m\sigma)$ labelling
its spin ($\sigma \! = \uparrow,\downarrow$) and orbital ($m=1,2,3$).
$\hat N_i =\sum_{\nu}\hat{d}^{\dagger}_{i\nu}\hat{d}^\pdag_{i\nu}$ is
the total number operator for site $i$ and $\hat{\mathbf S}_i$ its
total spin, with components $\hat S_i^\alpha =
\sum_{m\sigma\sigma'}\hat{d}^{\dagger}_{i m\sigma}
\tfrac{1}{2}\sigma^\alpha_{\sigma\sigma'}\hat{d}_{i m\sigma'}$, where
$\sigma^{\alpha}$ are Pauli matrices.  We study a Bethe lattice with
nearest-neighbor hopping amplitude $t$, used as energy unit
($t=1$). Onsite interactions are described by $\hat H_\interact$.
\todo{Kathi, to motivate this particular form, we need to add a
  reference!}  The onsite Coulomb interaction $U$ penalizes double
occupancy. The ferromagnetic coupling $J>0$ accounts for Hund's first
rule by favoring a large spin per site.  We choose the chemical
potential $\mu$ such that the filling per lattice site is one below
half-filling, $\langle \hat N_i \rangle \simeq 2$, inducive to an
intricate interplay of spin and orbital degrees of freedom.

\paragraph*{Methods.---}
\label{sec_Methods}
We use single-site DMFT to map the lattice model onto a 3-band
Anderson-Hund model (AHM) of the form $\hat H_\AHM = \varepsilon_d
\,\hat{N}+\hat{H}_\interact [\hat d^\dag_\nu] +\hat{H}_{\bath}$. Here
$d^\dag_\nu$ creates a local (``impurity'') electron of flavor $\nu$
with energy $\varepsilon_d=-\mu$, experiencing local interactions
$\hat H_\interact$, with total number and spin operators $\hat N$ and
$\hat {\mathbf{S}}$ defined analogously to $\hat N_i$ and $\hat
{\mathbf{S}}_i$.  The local site on average hosts two electrons ($n_d
= \langle \hat N \rangle \simeq 2$), forming a spin triplet and
orbital triplet (the one hole relative to half-filling can be in one
of three orbital levels).
The local electrons hybridize with a
3-band spinful bath, 
\begin{eqnarray}
\label{eq:hyb}
\hat H_\bath = 
\sum_{k\nu}\left(\varepsilon_{k\nu} 
c^{\dagger}_{k\nu}\hat{c}^\pdag_{k\nu} + 
V_{k}\bigl[\hat{d}^{\dagger}_{\nu}\hat{c}^\pdag_{k\nu} 
+ \hat{c}^{\dagger}_{k\nu} \hat{d}^\pdag_{\nu} \bigr] \right ) , \qquad 
\end{eqnarray}
with a hybridization function $\Gamma(\varepsilon) = \pi \sum_k
|V_k|^2 \delta(\varepsilon - \varepsilon_k)$ that fully characterizes
the impurity-bath interplay. In DMFT, $\Gamma(\varepsilon) $ has the
role of the effective Weiss mean field and is determined
self-consistently \cite{Georges1996,Kotliar2006,supplement}.
We studied both the self-consistent AHM (scAHM), and for comparison
also the pure impurity AHM (iAHM) without self-consistency, using a
flat density of states with half-bandwidth $D$, $\Gamma (\varepsilon)
\equiv\Gamma \Theta(D-|\varepsilon|)$.

We use full-density-matrix (fdm) NRG \cite{Weichselbaum2007}
exploiting non-Abelian symmetries \cite{Weichselbaum2012b}, both to
solve the iAHM and for each scAHM iteration (for NRG details, see
\cite{supplement}). The key idea of NRG, due to Wilson
\cite{Wilson1975}, is to discretize the bath's continuous spectrum
logarithmically, map the model onto a semi-infinite ``Wilson'' chain
with exponentially decaying hopping amplitudes, and exploit this
energy-scale separation to iteratively diagonalize the model while
discarding high-energy states.  This allows one to zoom in on
low-energy properties, at the expense of having only coarse-grained
resolution at high energies.
Nevertheless, NRG  results are accurate also for spectral integrals
even if these include large energies, since they can be evaluated
using discrete, unbroadened NRG data.

\paragraph*{Matsubara benchmark.---}
\label{sec_benchmark}

\begin{figure}
\centering
\includegraphics[width=1\linewidth,trim=0mm 3mm 0mm 0mm]{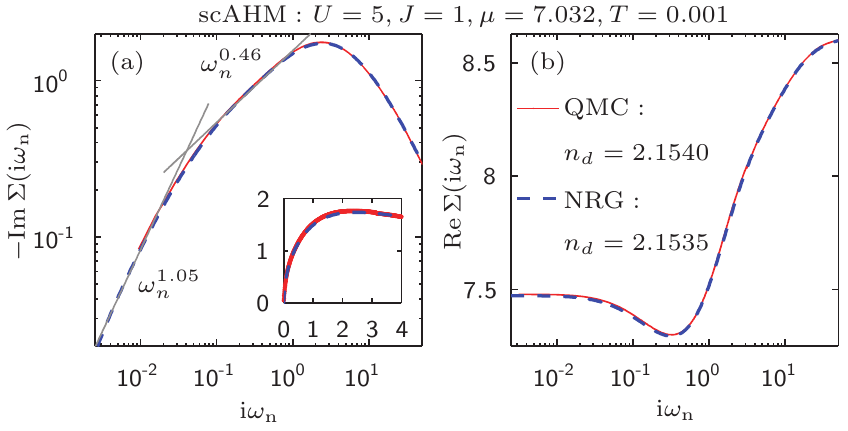}  
\caption{(Color online) Benchmark comparison of NRG and ctQMC for the
  3-band scAHM.  (a) Imaginary and (b) real part of the
  self-consistently converged self-energy as function of Matsubara
  frequencies. Grey lines in (a) are power-law fits to low and intermediate
  frequency data, respectively. The inset of (a) shows ${\rm Im}
  \,\Sigma({\rm i}\omega_n)$ on a linear scale.  \scrap{We used $N^{\rm
    max}_{\rm keep}=?$ within NRG. \todo{Kathi, please
remove the $i$ before $\omega_n$ on the x-Axis.}}
\todo{Put (a,b) next to each other, to save space. }}
\label{fig:benchmark} \vspace{-5mm}
\end{figure}
We illustrate this by benchmarking NRG versus ctQMC \cite{supplement},
which treats the bath as a continuum and has no bath discretization
issues.  We used both methods to compute the self-energy $\Sigma (\mi
\omega_n)$ of the Matsubara correlator $G(\mi\omega_n)$ associated
with the retarded local correlator $G^R(\omega) = \langle d^\pdag_\nu 
\mbox{$\parallel$}\, d_\nu^\dag \rangle_\omega$.  In NRG, its spectral function is
expressed in terms of discrete data, $A(\omega) = - \tfrac{1}{\pi}
{\rm Im} G^R(\omega)$ $\stackrel{{\scriptscriptstyle {\rm
      NRG}}}{\simeq} \sum_s a_s \delta(\omega - \xi_s)$, hence $G(\mi
\omega_n) = \int d\omega A(\omega)/(\mi \omega_n-\omega)
\stackrel{{\scriptscriptstyle {\rm NRG}}}{\simeq} \sum_s
a_s/(\mi\omega_n - \xi_s)$.  Fig.~\ref{fig:benchmark} compares NRG and
ctQMC results for $\Sigma (\mi \omega_n)$ at $T=0.001$. The agreement
is excellent, also at large frequencies.  The numerical costs differ
vastly, however, $\simeq 10^2$ vs.\ $10^{5}$ CPU hours
\cite{supplement}, since the chosen temperature is challengingly low
for ctQMC, whereas NRG can access any temperature.
Luttinger pinning at zero frequency
\cite{Muller-Hartmann1989,Georges1996} is fulfilled within 1.5$\%$ for
both methods \cite{supplement}.   ${\rm
  Im}\, \Sigma(\mi \omega_n)$ displays fractional power-law behavior for
intermediate frequencies ($0.1 \lesssim \omega_n \lesssim 1$), as
found in \cite{Werner2008,Yin2012}, and Fermi-liquid behavior for very
low frequencies, ${\rm Im} \,\Sigma({\rm i}\omega_n)\propto\omega_n$,
as found in \cite{Haule2009,Mravlje2011}.

\begin{figure}
\centering
\includegraphics[width=1\linewidth, trim=0mm 5mm 0mm
3mm]{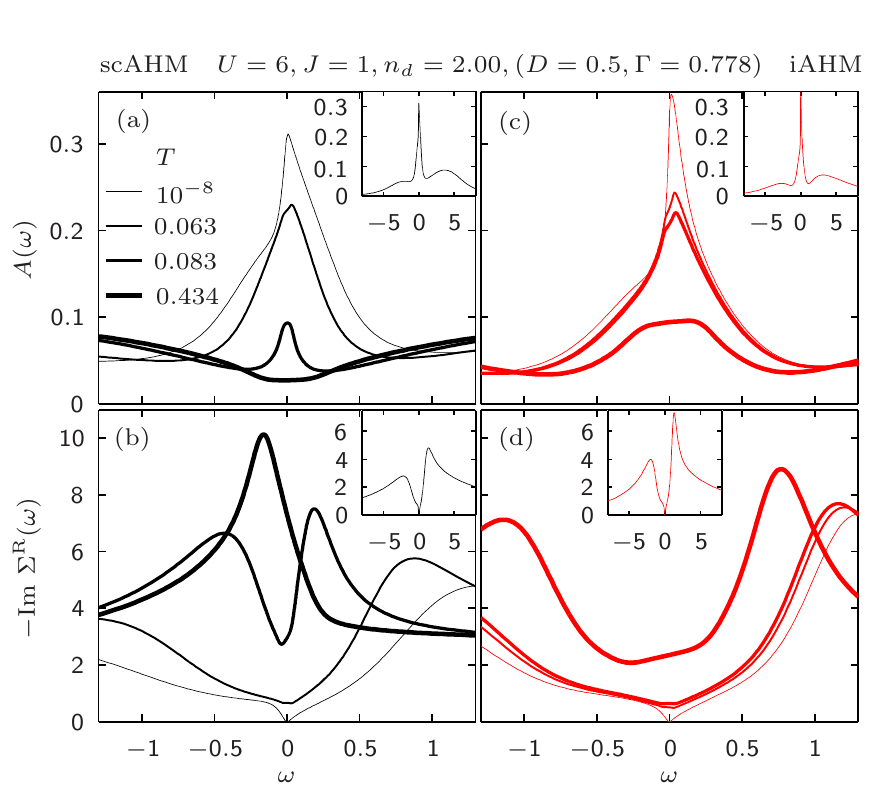}
\caption{(Color online) (a) The local spectral function $A(\omega)$
  and (b) the imaginary part of the retarded self-energy, ${\rm
    Im}\, \Sigma^R(\omega)$, for the scAHM, plotted versus frequency for
  four temperatures. 
 Insets show a larger   frequency range for
$T = 10^{-8}$. 
 (c,d) Same as in (a,b), but for an iAHM.}  
\label{fig:Tdep} \vspace{-6mm}
\end{figure}

\paragraph{Coherence-incoherence crossover.}
We now turn to real-frequency properties [Fig.\ \ref{fig:Tdep}].  At
zero temperature, the local spectral function $A( \omega)$ of the
scAHM shows a well-defined low-energy quasiparticle peak and $- {\rm
  Im}\, \Sigma^R(\omega)$ a dip reaching down to zero [insets of
Figs.~\ref{fig:Tdep}(a,b)].  This indicates that strong Kondo-type
screening correlations exist between bath and local spin and orbital
degrees of freedom.  At higher energies, $A(\omega)$ also shows
incoherent, rather flat particle-hole asymmetric side peaks, that
reflect charge fluctuations.  With increasing temperature, a
coherence-incoherence crossover occurs\tocite{}: the quasiparticle
peak weakens and eventually gives way to a pseudogap
[Fig.~\ref{fig:Tdep}(a)], and concurrently the dip in $- {\rm Im}\,
\Sigma^R(\omega)$ is smeared out into a broader minimum, which
eventually evolves into a maximum [Fig.~\ref{fig:Tdep}(b)].  During
this crossover, quasiparticle weight is transferred from low to high
energies, in a way reminiscent of recent photoemission measurements
\cite{Yi2013} (though the interpretation offered there invoked an
orbital-selective Mott phase).

The behavior seen in Figs.~\ref{fig:Tdep}(a,b) is characteristic of
``Hund's physics'', arising when the Hund's coupling $J$ is sizeable
($J/U$ nonzero and not too small). It contrasts markedly with the
behavior associated with a Mott transition, as exemplified, e.g., in
Figs.~2 and 3 of \cite{Bulla1999} for the 1-band Hubbard model:
the side peaks in our $A(\omega)$ are rather flat, the maxima in our $-{\rm Im}
\, \Sigma^R(\omega)$ never diverge, and the spectral weight near $\omega
\simeq 0$ remains nonzero at all temperatures, implying that metallic
behavior persists.  In this sense, Hund's metals exhibit properties
quite distinct from Mott-Hubbard systems. 

Since the scAHM is based on an impurity model, it is instructive to
study a corresponding iAHM, with parameters tuned to yield a similar
spectral function at $T=0$ [Fig.~\ref{fig:Tdep}(c,d)]. It likewise
features a large low-energy (Kondo) peak that weakens with increasing
temperature, though no pseudogap occurs. This shows that the $T=0$
spectral properties of the scAHM are governed by the impurity physics
of the iAHM, while the scAHM transfer of spectral weight with increasing $T$
is brought about by DMFT self-consistency.

\begin{figure}
\includegraphics[width=\linewidth, trim=0mm 80mm 0mm 12mm]{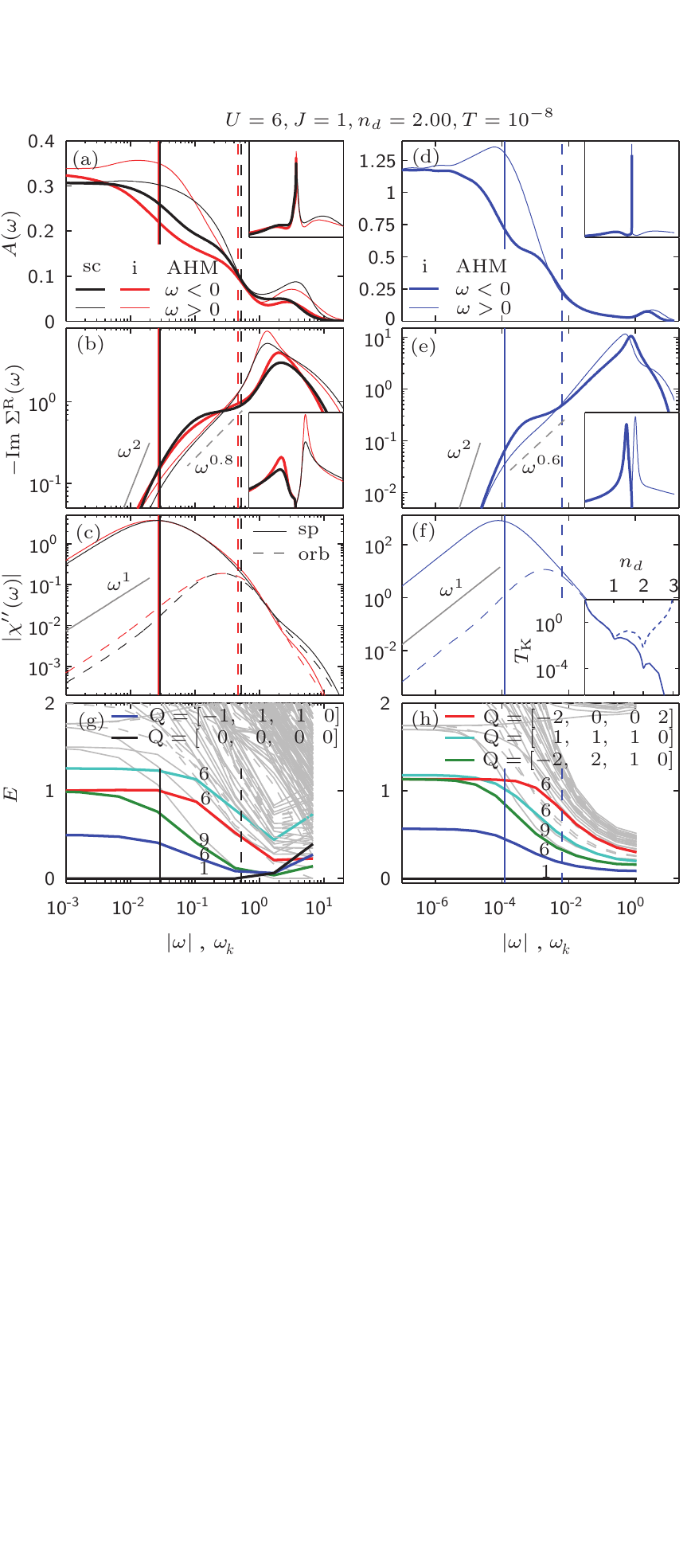}
\caption{(Color online) (a-f) Spin-orbital separation in
  real-frequency, ground state correlators. 
  The left column uses the same parameters and
  color code as Fig.~\ref{fig:Tdep} for the scAHM (black) and iAHM
  with $(\Gamma,D) = (0.778,0.5)$ (red). For comparison, the right
  column shows iAHM results with $(\Gamma,D)= (0.200,1.0)$ (blue),
  yielding smaller crossover scales.  (a,d) The local spectral
  functions, (b,e) the local self-energy, and (c,f) the spin and
  orbital susceptibilities, $\chi''_\spin$ (solid) and $\chi''_\orb$
  (dashed). We use a logarithmic frequency scale, with thick (thin)
  lines for $\omega<0$ ($\omega>0$).  Insets show data on a linear
  scale.  In all panels, solid (dashed) vertical lines mark the spin
  (orbital) Kondo scale, $\Tkspin$ ($\Tkorb$).  Grey guide-to-the eye
  lines indicate Fermi-liquid power laws (solid) 
or apparent fractional   power laws (dashed).
  Inset to (f): Kondo scales $\Tkspin$ (solid) and $\Tkorb$ (dashed)
  for the iAHM, plotted as function of $n_d$.  (g,h) NRG eigenlevel
  flow diagrams for the scAHM and iAHM of panels (a-c) and (b-d),
  respectively, showing the rescaled energies of the lowest-lying
  eigenmultiplets of a Wilson chain of (even) length $k$, plotted
  versus its characteristic level spacing $\omega_k =
  \Lambda^{-k/2}$ (see text).  Numbers above lines give multiplet degeneracies, $Q$ their symmetry labels. \vspace{-6mm} 
}
\label{fig:DMFTvsimp}
\end{figure}

\paragraph*{Particle-hole asymmetry.}
We next exploit the power of NRG to zoom in to arbitrarily low energy
scales: In Figs.~\ref{fig:DMFTvsimp}(a,b), we replot on a logarithmic
scale the data (black/red for sc/iAHM) from Figs.~\ref{fig:Tdep}(a-d)
for $A(\omega)$ and ${\rm Im}\, \Sigma^R(\omega)$ at $T=10^{-8}$.  For
comparison, the right column of Fig.~\ref{fig:DMFTvsimp} again shows
results for the iAHM, but using parameters that yield smaller
crossover scales (defined below), to better separate the
low-energy features associated with spin and orbital screening from
high-energy features associated with charge fluctuations.  Note again
the striking qualitative similarity between the scAHM (black) and iAHM
(red/blue) spectra -- clearly, for $T\simeq 0 $ DMFT self-consistency
plays no major role. 

With decreasing temperature, the quasiparticle peaks in
Figs.~\ref{fig:Tdep}(a,c) show an increasing particle-hole asymmetry,
not surprising away from half-filling, which at $T\simeq 0$ is very
pronounced: in Figs.~\ref{fig:DMFTvsimp}(a,d) for $A(\omega)$, the
thick ($\omega < 0$) lines show a shoulder-like structure for
intermediate frequencies (between the vertical solid and dashed
lines), while the thin ($\omega > 0$) lines do not; and in
Figs.~\ref{fig:DMFTvsimp}(b,e) for ${\rm Im}\, \Sigma^R(\omega)$, the
thick lines show a plateau-like structure, whereas the thin lines show
approximate $\sim \omega^\alpha$ power-law behavior (with
non-universal $\alpha$).  For the Matsubara self-energy obtained via
the Hilbert transform $\Sigma (\mi \omega_n) = - \frac{1}{\pi} \int d
\omega {\rm Im} \, \Sigma^R(\omega)/(\mi \omega_n - \omega)$, the
asymmetric contributions from the power-law and shoulder in ${\rm Im}
\, \Sigma^R(\omega \stackrel{\scriptscriptstyle >}{\scriptscriptstyle
  <}0)$ conspire in such a way that ${\rm Im}\, \Sigma (\mi \omega_n)$
shows an apparent fractional power law, but ${\rm Re}\, \Sigma (\mi
\omega_n)$ does not [Fig.~\ref{fig:benchmark}(a,b)]. Conversely, this
example ilustrates that care is due when drawing real-frequency
conclusions from imaginary-frequency power laws: if one is present
only for ${\rm Im}\, \Sigma (\mi \omega_n)$, not ${\rm Re}\, \Sigma
(\mi \omega_n)$ (as in \cite{Werner2008,Yin2012}), then ${\rm Im}\,
\Sigma^R(\omega)$ need not show pure power-law behavior.

\paragraph*{Spin-orbital separation.}
Next we elucidate the screening of local spin and orbital degrees of
freedom by the bath of conduction electrons. To this end,
Figs.~\ref{fig:DMFTvsimp}(c,f) respectively show the imaginary part
$(\chi'')$ of the dynamical spin and orbital susceptibility,
$\chi_\spin = \sum_\alpha \langle \hat S^\alpha \mbox{$\parallel$}
\hat S^\alpha \rangle_\omega$ and $\chi_\orb = \sum_a \langle \hat T^a
\mbox{$\parallel$} \hat T^a \rangle_\omega$, with orbital operators
$\hat T^a = \sum_{mm'\sigma}\hat{d}^{\dagger}_{m\sigma} \tfrac{1}{2}
\tau^a_{mm'}\hat{d}_{m'\sigma}$, where $\tau^a$ are the SU(3)
Gell-Mann matrices, normalized as ${\rm Tr}[ \tau^a\tau^b ]
=2\delta_{ab}$.  Both $\chi''_\spin$ and $\chi''_\orb$ exhibit a peak
with (nearly) power-law flanks, characteristic of Kondo screening of
the local spin and orbital degrees of freedom. Strikingly, for both
scAHM and iAHM the peak for $\chi''_\spin$ occurs at a much lower
energy and is much higher than for $\chi''_\orb$.  The peak positions
lie near the spin and orbital Kondo scales, $\Tkspin$ and $\Tkorb$
(vertical solid and dashed lines), calculated from
$\chi_\spin(\omega)$ and $\chi_\orb(\omega)$ using the recipe of
\cite{Hanl2013}. $\Tkspin$ acts as the coherence scale below which
Fermi-liquid behavior (${\rm Im} \Sigma^R(\omega) \propto \omega^2$,
$\chi''_{\spin, \orb} \propto \omega$, see
Figs.~\ref{fig:DMFTvsimp}(b-f), grey lines) sets in.  The
SU(2)${}_\spin$ and SU(3)${}_\orb$ crossover scales differ strongly,
$\Tkspin \ll \Tkorb$, because the Kondo temperature for an SU($N$)
Kondo model scales as $\ln \Tk \sim - 1/N$.  This implies two-stage
screening, with spin screening occuring at significantly lower
energies than orbital screening.  This ``spin-orbital separation'',
featuring a very small coherence scale and an intermediate regime with
screened orbital degrees of freedom coupled to slowly fluctuating,
large spins, is a central result of our work.

The inset of Fig.~\ref{fig:DMFTvsimp}(f) depicts $\Tkspin$ and
$\Tkorb$ for the iAHM as function of the filling $n_d$.  For $n_d
\simeq 1$, where the bare impurity's ground state has SU(6) symmetry
also for $J \neq 0$, $\Tkspin \simeq \Tkorb\simeq \Tk^{\rm SU(6)}$. As
$n_d$ increases from 1 to 2, $\Tkspin$ and $\Tkorb$ split apart and
spin-orbital separation sets in.  As $n_d$ continues to increase
towards 3, $\Tkorb$ becomes very large ($\gtrsim D$), reflecting the
fact that for half-filling the orbitals form an orbital singlet from
the outset. In this sense, $n_d \simeq 2$ is special: there conditions
are optimal for the Hund coupling to align two spins in different
orbitals without forming an orbital singlet.

\paragraph*{RG flow.}
In renormalization group (RG) terms, the two-stage screening discussed
above is associated with the RG flow between three fixed points,
describing \mbox{high-}, intermediate- and low-energy energy
excitations. Their effective fixed point Hamiltonians have ground
state multiplets whose spin$\times$orbit structure is
triplet$\times$triplet, triplet$\times$singlet and
singlet$\times$singlet, implying a ground state entropy of $\ln(9)$,
$\ln(3)$ and $\ln(1)$, respectively.  The RG flow between these fixed
points can be visualized via NRG eigenlevel flow diagrams
[Fig.~\ref{fig:DMFTvsimp}(g,h)]. Technically, they show how the
lowest-lying rescaled eigenlevels of a length-$k$ Wilson chain evolve
with $k$, where ``rescaled'' means given in units of $\omega_k =
\Lambda^{-k/2}$ ($\Lambda >1$ being a discretization parameter).
Conceptually, these levels represent the finite-size spectrum of the
impurity+bath put in a spherical box of radius $R_k \sim
\Lambda^{k/2}$, centered on the impurity \cite{Wilson1975,Delft1998}:
as $k$ increases, the finite-size level spacing $\omega_k \propto
1/R_k$ decreases exponentially. The structure of the finite-size
spectrum is fixed ($k$-independent) while $\omega_k$ lies within an
energy regime governed by one of the fixed points, but changes when
$\omega_k$ passes through a crossover scale between two fixed points.

Figs.~\ref{fig:DMFTvsimp}(g,h) show this RG flow for the scAHM and the
iAHM, revealing similar behavior for both \cite{details-differ}.  We
label multiplets by their $\text{U(1)}_\charge\times%
\text{SU(2)}_\spin \times\text{SU(3)}_\orb$ symmetry labels,
$Q=[q,2S,q_1q_2]$; here $q$ denotes particle number relative to
half-filling, $S$ spin, and $(q_1q_2)$ an SU(3) irrep, identified by a
Young diagram with $q_1+q_2$ ($q_2$) boxes in its first (second) row.
The flow of the lowest-lying levels reveals two crossover scales,
$\Tkorb$ and $\Tkspin$ (whose spacing, though, is too small for the
level flow in between to become stationary, i.e.\ $k$-independent).
As $\omega_k$ drops below $\Tkorb$, orbital screening sets in,
favoring orbital singlets [$(q_1q_2) = (00)$], hence other multiplets
rise in energy, the more so the larger $(q_1q_2)$. Similarly, as
$\omega_k$ drops below $\Tkspin$, spin screening sets in, favoring
spin singlets and pushing up multiplets with $S \neq 0$.  For
$\omega_k \ll \Tkspin$, the ground state is a spin and orbital singlet
[$Q=(0,0,00)$]. We have checked that its excitation spectrum can be
interpreted in terms of non-interacting single-particle excitations,
thus confirming its Fermi-liquid nature.

\paragraph*{Conclusions.}
We have demonstrated the potential of DMFT+NRG as real-frequency
method to treat multi-orbital systems, with no need for analytic
continuation.  Applied to a Hund's metal, it revealed subtle spectral
features which are manifestly different from those of Mott-Hubbard
systems, and which can be probed in photoemission and STM
spectroscopies.  Our work is a first step towards using LDA+DMFT+NRG
to calculate AC and DC transport properties in
strongly correlated materials. Such applications will typically
involve less orbital symmetries than the model studied here, but could
be treated using the recent ``interleaved'' NRG approach of
\cite{Mitchell2014}. The latter yields results of comparable accuracy
and efficiency as when symmetries can be exploited \cite{Stadler2015},
and calculations with up to 5 bands seem within reach.

A key advantage of NRG is its ability to iteratively uncover the
system's RG flow from high to low energies, revealing the relevant
physics at each scale.  In the context of Mott-Hubbard systems, RG
ideas have been very fruitful even in very approximate implementations
\cite{Moeller1995,Fisher1995,Held2013}.  For the present Hund's metal,
the numerically exact RG flow achieved via DMFT+NRG revealed a clear,
simple picture of the crossover from the incoherent to the coherent
Fermi-liquid regime: two-stage screening of first orbital, then spin
degrees of freedom. Using DMFT+NRG to gain this type of RG understanding
of real material properties would be a worthwhile goal for future
research.

\acknowledgements
We acknowledge fruitful discussions with 
C.\ Aron, A.\ Georges, K.\ Haule, A.\ Mitchell and G.\ Zar\'and.
KMS, AW and JvD were supported by DFG (SFB-TR12, SFB631,
NIM, WE4819/1-1 and WE4819/2-1), ZPY and GK by NSF DMR1308141.



\begin{thebibliography}{52}
\bibitem [{det()}]{details-differ}%
  \BibitemOpen
  \href@noop {} {\bibinfo  {journal} {Details differ for $\omega_k \gtrsim
  \Tkorb$, because the scAHM crossover scales on the left are larger than on
  the right\cite{aronmodel}.}\ }\BibitemShut {NoStop}%
\end{thebibliography}

%

\section*{Supplementary Material}

\paragraph*{Recent NRG progress.}

While Wilson's NRG \cite{Wilson1975,Bulla2008} has been tremendously
successful in the past, significant further progress was achieved in
recent years, triggered by the realizations that it (i) can be
formulated in MPS language \cite{Weichselbaum2005,Weichselbaum2012a},
and (ii) that the discarded states can be used to construct a complete
many-body basis of approximate energy eigenstates \cite{Anders2005}.
It has become possible (iii) to reliably calculate finite-temperature
spectral functions in sum-rule conserving fashion
\cite{Peters2006,Weichselbaum2007}\comment{(sum-rule conserving, but
  only defined at T=0?)}; \todo{yielding a discrete representation of
  the form $A(\omega) = \sum_s a_s \delta(\omega - \xi_s)$, with
  $\sum_s a_s = 1$.} (iv) to treat a bath with nontrivial
hybridization function by suitably optimizing its representation in
terms of discrete bath states
\cite{Zitko2009,Zitko2009a,Stadler2015a}; to treat multi-band models,
either (v) by exploiting non-Abelian symmetries
\cite{Toth2008,Weichselbaum2012b} if the bands couple symmetrically to
the impurity \cite{Moca2012,Hanl2013,Hanl2014}, as here, or (vi) by
using an ``interleaved'' discretization scheme
\cite{Mitchell2014,Stadler2015}; and (vii) to greatly increase
numerical efficiency by optimizing MPS bond dimensions and to ensure
accuracy by checking the discarded weight
\cite{Weichselbaum2011a}. Taken together, these advances make NRG a
highly competitive real-frequency impurity solver for DMFT.

\paragraph*{Details of our NRG implementation.}

Wilson's NRG approach is based on logarithmically coarse-graining the
continuous bath in energy space into intervals of exponentially
decreasing widths in order to resolve even the lowest relevant energy
scale of the impurity system.  To accurately represent the
frequency-dependent hybridization function of the DMFT
self-consistency condition in terms of a set of discrete states, we
use a numerically stable implementation \cite{Stadler2015a} of the
discretization scheme of Ref.~\cite{Zitko2009,Zitko2009a}. 

The model is then mapped onto a semi-infinite ``Wilson'' chain with
exponentially decaying hopping amplitudes. This energy-scale
separation is exploited to iteratively diagonalize the model by adding
one site at a time, while discarding high-energy states.  The accuracy
of this truncation procedure can be checked by estimating the
discarded weight $\delta\rho_{\rm disc}$ \cite{Weichselbaum2011a}, a
quantitative convergence criterion inspired by DMRG, after each run.
Empirically, an NRG run is well-converged when $\delta \rho
_\mathrm{disc} < 10^{-12}$.

We exploit all available non-Ablian symmetries of the AHM studied
here, namely $\text{U(1)}_\charge\times%
\text{SU(2)}_\spin \times\text{SU(3)}_\orb$, by using the QSpace
approach developed by A. Weichselbaum \cite{Weichselbaum2012b}.
QSpace is a tensor library that is able to treat Abelian and
non-Abelian symmetries on a generic level.  In the presence of
symmetries, the state space can be organized into symmetry multiplets,
and tensors ``factorize'' into two parts, acting in the reduced
multiplet space and the Clebsch Gordon coefficient space,
respectively, vastly reducing numerical cost.  

We use full-density-matrix (fdm)NRG
\cite{Weichselbaum2007,Weichselbaum2012a} to calculate high-quality,
sum-rule-conserving spectral functions at arbitrary temperature.  To
implement the DMFT self-consistency loop (described below) we smoothen
the discrete spectral data provided by fdmNRG using the log-Gaussian
smoothening approach of Ref.~\cite{Weichselbaum2007}.  To improve the
resolution of the smoothened spectral data we calculate the
self-energy in every iteration from the ratio of two correlation
functions \cite{Bulla1998}. Moreover, for the last DMFT iteration we
average over two versions of the discretization grid ($z$-averaging
with $N_z = 2$, \cite{Oliveira1994,Zitko2009,Zitko2009a}). We have
explored also doing $z$-averaging for earlier DMFT-iterations, but
that did not noticably improve convergence.

\paragraph*{NRG-related computational parameters.}
\label{sec_NRG-parameters}

The performance of NRG is governed by the following computational
parameters: the dimensionless logarithmic discretization parameter
$\Lambda$; the truncation energy $E_{\rm trunc}$ in rescaled units (as
defined in Ref.~\onlinecite{Weichselbaum2012a}), up to which all
eigenmultiplets are kept, unless their number exceeds $N^{\rm
  max}_{\rm keep}$, the maximal number of kept multiplets per
iteration; the number $N_z$ of z-shifts for z-averaging
\tocite{Zitko2009,Zitko2009a,Oliveira1994}; and the log-Gaussian
broadening parameter $\sigma$ for smooth spectral data
\cite{Weichselbaum2007} .

We use $\Lambda=4$, $\sigma=0.8$ and $E_{\rm trunc}=7$.  For impurity
spectral functions and self-energies we used $N^{\rm max}_{\rm keep} =
2500$, yielding discarded weights of $\delta\rho_{\rm disc}<10^{-11}$.
For the susceptibilities $\chi_\spin$ and $\chi_\orb$ we use smaller
value for $N^{\rm max}_{\rm keep} = 1500$ (because the operators
involved are numerically rather costly); this yields values for
$\Tkspin$ and $\Tkorb$ that are consistent with the crossover scales
derived from NRG eigenlevel flow diagrams.

\paragraph*{DMFT self-consistency loop.---}
\label{sec_dmft-selfconsistency}
In single-site DMFT a quantum lattice model is treated in a quantum
mean-field fashion. Spatial correlations are frozen out by dropping
the momentum-dependence of the lattice self-energy, whereas temporal
quantum fluctuations are retained\tocite{Metzner1989,?}.
The lattice dynamics is then fully captured by the local retarded
lattice Green's function, $G^R_{\rm {latt}}(\omega)$, given in terms of
the purely local but still frequency dependent self-energy,
$\Sigma^R (\omega),$ - or equivalently by the retarded Green's function
of an impurity model $G^R_{\rm imp}(\omega)$ with equal local
interactions (equal $\Sigma^R(\omega)$) and effective hybridization
$\Gamma(\omega)$.  This equivalence, $G^R_{\rm {latt}}(\omega)= G_{\rm
  imp}(\omega)\equiv G(\omega)$, imposes a self-concistency condition,
that simplifies to $\Gamma(\omega)=-t^2 {\rm Im}\, G^R_{\rm imp}(\omega)$
for a Bethe lattice, and fully maps the quantum lattice problem onto
an effective quantum impurity problem by iteratively determining
$\Gamma(\omega)$ \cite{Georges1996}.  In each
iteration of the DMFT self-consistent mapping, we solve the quantum
impurity model with NRG.

\paragraph*{Computational costs.}
Our ctQMC solver, developed by K. Haule, is based on an expansion in
the hybridization function \cite{Werner2006,Millis2006}.  Its
implementation is described in \cite{Haule2007}.  For the benchmark
calculations shown in Fig.~1, our ctQMC solver was run in parallel
using 2500 cores with $10^5$ Monte Carlo steps in each core.  To
achieve DMFT self-consistency for $T=0.001$, it needed 20 iterations
of 6 hours each.  Our NRG code, developed by A. Weichselbaum
\cite{Weichselbaum2012a,Weichselbaum2012b}, achieved DMFT
self-consistency after 7 iterations of 10 hours each, on a single
8-core machine with 128 GB memory. This results in nearly three orders
of magnitude better numerical efficiency. Moreover, NRG is able to
access arbitrarily low temperatures, whereas for ctQMC the numerical
costs grow exponentially with decreasing temperature.

\end{document}